# On Regulating Downstream AI Developers

Sophie Williams[1],✉   Jonas Schuett[1]   Markus Anderljung[1]

## Abstract

Foundation models – models trained on broad data that can be adapted to a wide range of downstream tasks – can pose significant risks, ranging from intimate image abuse, cyberattacks, to bioterrorism. To reduce these risks, policymakers are starting to impose obligations on the developers of these models. However, downstream developers – actors who fine-tune or otherwise modify foundational models – can create or amplify risks by improving a model's capabilities or compromising its safety features. This can make rules on upstream developers ineffective. One way to address this issue could be to impose direct obligations on downstream developers. However, since downstream developers are numerous, diverse, and rapidly growing in number, such direct regulation may be both practically challenging and stifling to innovation. A different approach would be to require upstream developers to mitigate downstream modification risks (e.g. by restricting what modifications can be made). Another approach would be to use alternative policy tools (e.g. clarifying how existing tort law applies to downstream developers or issuing voluntary guidance to help mitigate downstream modification risks). We expect that regulation on upstream developers to mitigate downstream modification risks will be necessary. Although further work is needed, regulation of downstream developers may also be warranted where they retain the ability to increase risk to an unacceptable level.

[1] Centre for the Governance of AI, Oxford, UK.
✉ Contact: sophie.williams@governance.ai.

# Table of contents

# Executive summary

*What is a downstream AI developer?* ([Section II.1](#))

- A downstream developer is any actor that modifies another actor's foundation model via fine-tuning, scaffolding, or other techniques.
- Downstream developers often tailor a model for a specific use case, task, or domain (e.g. for use in healthcare), but they can also make more general modifications.
- Downstream developers play a key role in delivering the potential benefits of foundation models to society, because the upstream developers that build them are unlikely to have the capacity or sector-specific knowledge to adapt them in many different ways.
- There are many downstream developers spread across the globe – from individuals through to large corporations – and this number is likely to increase.

*How can downstream developers increase risks?* ([Section II.2](#))

- Downstream developers can increase risks either by improving a foundation model's capabilities, or by compromising the safety features put in place by the upstream developer. This increase in risk can be intentional (e.g. if the model's safeguards are deliberately evaded) or unintentional (e.g. if the model's capabilities are improved in a legitimate way that can also be used to cause harm).
- The potential risks from modifications may be even greater if the original model is highly capable or if the model's weights are released.

*What challenges do downstream developers present to policymakers?* ([Section II.3](#))

- Downstream modifications that increase risk could undermine emerging regulatory frameworks that would place obligations on upstream developers to mitigate the risks from their models.
- Downstream developers may be unable to effectively identify and mitigate the risks from the modifications they have made if they lack sufficient access or information about the underlying model. On the other hand, upstream developers are unlikely to be able to predict or mitigate the risks from every possible downstream modification.
- The sheer number and diversity of downstream developers means that a one-size-fits-all approach is unlikely to be effective.
- Policymakers may also be concerned that any interventions to mitigate downstream modification risks may stifle innovation, especially as regulation can have a disproportionate impact on smaller actors.

*How might policymakers address risks posed by downstream developers?* ([Section III](#))

- One option would be to place obligations directly on downstream developers ("regulate downstream"). This raises questions about which downstream developers should be in scope, and what obligations should apply to them. Criteria could be used to restrict regulations to a subset of downstream developers (ideally, those likely to increase risk to an unacceptable level). Obligations could include, for example, requiring downstream developers to test the modified model for new dangerous capabilities so that additional mitigations can be put in place, or requiring them to report risks and incidents resulting from it. Any regulation would need to be carefully designed to avoid placing unnecessary burdens on downstream developers and enable those in scope to effectively identify and mitigate risks.
- A second option would be to place obligations on upstream developers to mitigate downstream modification risks ("regulate upstream"). This could include, for example, requiring upstream developers to put certain safeguards in place and limit certain downstream modifications. This may be appealing to policymakers aiming to target regulation at developers of the most powerful models. However, it is unlikely to address all potential risks stemming from downstream modifications and may leave the regulator unable to directly intervene in cases where upstream mitigations are ineffective. It may also risk stifling innovation if it encourages upstream developers to be overly restrictive.
- A third option would be to not introduce any new regulation with regards to downstream developers, but instead clarify how tort law applies to downstream developers or issue voluntary guidance to help mitigate downstream modification risks. This would have the benefit of keeping any regulatory regime narrow in scope, but may be insufficient in addressing the risks from downstream modifications.

*What actions should policymakers take?* ([Section IV](#))

- Policymakers should issue voluntary guidance for both upstream and downstream developers, setting out best practices for mitigating the risks from downstream modifications.
- When regulations on upstream developers are introduced, they should include some requirements to mitigate the risks from downstream modifications.
- Policymakers should also monitor the downstream developer landscape and be prepared to introduce targeted and proportionate obligations in cases where they retain the ability to increase risk to unacceptable levels. We offer a sketch for what such obligations could be and the conditions under which they could apply.

# I. Introduction

Foundation models – models trained on broad data that can be adapted to a wide range of downstream tasks[1] – can pose significant risks.[2] For example, they can be misused to create non-consensual deepfake pornography or child sexual abuse material.[3] More capable foundation models could enhance cybercriminals' ability to identify software vulnerabilities[4] or terrorists' ability to acquire and deploy biological weapons.[5] Even well-intentioned users might cause accidental harm in high-risk contexts if they lack sufficient information about the models they are using.

To mitigate these risks, regulators are starting to impose requirements on developers of foundation models. For example, the EU AI Act[6] already sets out obligations for providers of general-purpose AI models.[7] Among other things, providers will need to make information and documentation about their models available to certain users.[8] Providers of general-purpose AI models with systemic risk will also be required to, for example, perform model evaluations and ensure an adequate level of cybersecurity protection.[9] Some of these obligations are further concretised in a Code of Practice.[10] Similarly, an upcoming bill in the UK is expected to introduce obligations for developers of "the most powerful AI models",[11] alongside more targeted measures to combat AI-generated child sexual abuse material.[12] Concentrating regulatory efforts on "upstream" developers of foundation models is a sensible starting point because they are generally well-resourced and can take effective measures to mitigate risks.[13]

---

[1] R Bommasani et al., "On the Opportunities and Risks of Foundation Models" (arXiv, 2022) <http://arxiv.org/abs/2108.07258>.

[2] For an overview of different risks, see Y Bengio et al., "International AI Safety Report" (January 2025) <https://perma.cc/9JGF-B72D>; OECD, "Assessing Potential Future Artificial Intelligence: Risks, Benefits and Policy Imperatives" (November 2024) <https://perma.cc/P3US-UPHA>.

[3] See D Harris, "Deepfakes: False Pornography Is Here and the Law Cannot Protect You" (2019) 17 Duke Law & Technology Review 99; M Westerlund, "The Emergence of Deepfake Technology: A Review" (2019) 9 Technology Innovation Management Review 40 <http://doi.org/10.22215/timreview/1282>; OpenAI, "OpenAI's Commitment to Child Safety: Adopting Safety by Design Principles" (2024) <https://openai.com/index/child-safety-adopting-sbd-principles> accessed 10 March 2025.

[4] See M Brundage et al., "The Malicious Use of Artificial Intelligence: Forecasting, Prevention, and Mitigation" (arXiv, 2018) <http://arxiv.org/abs/1802.07228>; B Guembe et al., "The Emerging Threat of AI-Driven Cyber Attacks: A Review" (2022) 36 Applied Artificial Intelligence 2037254 <https://doi.org/10.1080/08839514.2022.2037254>; AJ Lohn and KA Jackson, "Will AI Make Cyber Swords or Shields?" (Center for Security and Emerging Technology 2022) <https://doi.org/10.51593/2022CA002>.

[5] See JB Sandbrink, "Artificial Intelligence and Biological Misuse: Differentiating Risks of Language Models and Biological Design Tools" (arXiv, 2023) <http://arxiv.org/abs/2306.13952>; EH Soice et al., "Can Large Language Models Democratize Access to Dual-Use Biotechnology?" (arXiv, 2023) <http://arxiv.org/abs/2306.03809>; DA Boiko, R MacKnight and G Gomes, "Emergent Autonomous Scientific Research Capabilities of Large Language Models" (arXiv, 2023) <http://arxiv.org/abs/2304.05332>; A Gopal et al., "Will Releasing the Weights of Future Large Language Models Grant Widespread Access to Pandemic Agents?" (arXiv, 2023) <https://arxiv.org/abs/2310.18233>; RF Service, "Could Chatbots Help Devise the Next Pandemic Virus?" (2023) 380 Science 1211 <https://doi.org/10.1126/science.adj2463>.

[6] Regulation (EU) 2024/1689 Laying Down Harmonised Rules on Artificial Intelligence (Artificial Intelligence Act) [2024] OJ L 2024/1689 <https://eur-lex.europa.eu/eli/reg/2024/1689/oj/eng>.

[7] See Chapter V, in particular Art. 53, 55.

Much less attention has been paid to the regulation of downstream developers[14] – actors who fine-tune or otherwise modify foundational models. This could leave a major gap in regulatory frameworks because downstream developers could make modifications that undermine safety measures taken upstream.[15] For example, an upstream developer might train their model to refuse

---

[8] Art. 53(1)(b).

[9] Art. 55(1)(a), (d).

[10] European Commission, "General-Purpose AI Code of Practice" (2024) <https://perma.cc/CQK4-HN9X>.

[11] HM King Charles III, "The King's Speech 2024" (17 July 2024) <https://perma.cc/3LH5-7M5G>.

[12] The UK Crime and Policing Bill will introduce a new offence which criminalises AI models that have been optimised to create child sexual abuse material, see UK Home Office and UK Ministry of Justice, "Crime and Policing Bill: Child Sexual Abuse Material Factsheet" (15 February 2025) <https://perma.cc/4S9L-LTF9>.

[13] E.g. many upstream developers have implemented specific risk management frameworks, e.g. Anthropic, "Responsible Scaling Policy" (15 October 2024) <https://perma.cc/5GK4-W3KH>; OpenAI, "Preparedness Framework (Beta)" (18 December 2023) <https://perma.cc/MBP2-SL6C>; Google DeepMind, "Frontier Safety Framework (Version 2.0)" (4 February 2025) <https://perma.cc/28ZW-D8K4>.

[14] Downstream developers already need to navigate a complex legal framework. For example, the EU General Data Protection Regulation (GDPR) applies to the processing of personal data used to train or fine-tune a foundation model (Regulation [EU] 2016/679 On the Protection of Natural Persons With Regard to the Processing of Personal Data and on the Free Movement of Such Data [General Data Protection Regulation] OJ L 119/1 <https://eur-lex.europa.eu/eli/reg/2016/679/oj/eng>) and the EU Cyber Resilience Act (CRA) imposes cybersecurity-by-design standards on manufacturers of products with digital elements, which also includes AI systems (Regulation [EU] 2024/2847 On Horizontal Cybersecurity Requirements for Products With Digital Elements [Cyber Resilience Act] OJ L 2024/2847 <https://eur-lex.europa.eu/eli/reg/2024/2847/oj/eng>).

[15] See C Anil et al., "Many-Shot Jailbreaking" (Advances in Neural Information Processing Systems, San Diego, 2024) <https://perma.cc/H9L3-ALAP>; X Qi et al., "Fine-Tuning Aligned Language Models Compromises Safety, Even When Users Do Not Intend To!" (arXiv, 2023) <http://arxiv.org/abs/2310.03693>; T Davidson et al., "AI Capabilities Can

requests for advice on how to acquire chemical weapons, but a downstream developer could potentially override that training. Policymakers could choose to accept this potential loophole, relying on existing rules to try to address downstream modification risks. Alternatively, they could introduce rules designed to close it. They could either impose additional requirements on upstream developers or regulate downstream developers directly.[16] However, each approach poses its own set of challenges. Notably, direct regulation of downstream developers could significantly expand the scope of the regime.

This limited focus on downstream developers is also reflected in the literature on foundation model regulation. Although there has been some work on the AI value chain,[17] existing research on regulation has mainly focused on upstream developers.[18] However, there is a growing body of technical work illustrating how downstream developers can increase the risks posed by foundation models, either through modifications that improve their capabilities[19] or

---

Be Significantly Improved without Expensive Retraining" (arXiv, 2023) <http://arxiv.org/abs/2312.07413>; P Gade et al., "BadLlama: Cheaply Removing Safety Fine-Tuning from Llama 2-Chat 13B" (arXiv, 2024) <http://arxiv.org/abs/2311.00117>; X Yang et al., "Shadow Alignment: The Ease of Subverting Safely-Aligned Language Models" (arXiv, 2023) <http://arxiv.org/abs/2310.02949>; D Halawi et al., "Covert Malicious Fine-tuning: Challenges in Safeguarding LLM Adaptation" (arXiv, 2024) <http://arxiv.org/abs/2406.20053>.

[16] Note that the EU AI Act acknowledges that general-purpose AI models can be "modified or fine-tuned into new models" (Recital 97). This implies that downstream developers could be subject to obligations that apply to general-purpose AI model providers. However, more clarity is needed about the circumstances under which this would be the case. The EU AI Office has indicated that they expect very few downstream developers to be in scope.

[17] See I Brown, "Allocating Accountability in AI Supply Chains: A UK-Centred Regulatory Perspective" (Ada Lovelace Institute 2023) <https://perma.cc/HTB7-BVBN>; S Küspert, N Moës and C Dunlop, "The Value Chain of General-Purpose AI" (Ada Lovelace Institute 2023) <https://perma.cc/JD3V-TALN>; UK Competition & Markets Authority, "AI Foundation Models: Technical Update Report" (2024) <https://perma.cc/SXP5-KB5A>.

[18] E.g. P Hacker, A Engel and M Mauer, "Regulating ChatGPT and Other Large Generative AI Models," (ACM Conference on Fairness, Accountability, and Transparency, Chicago, 2023) <https://doi.org/10.1145/3593013.3594067>; J Schuett et al., "From Principles to Rules: A Regulatory Approach for Frontier AI" in P Hacker, A Engel, S Hammer and B Mittelstadt (eds), *The Oxford Handbook on the Foundations and Regulation of Generative AI* (Oxford University Press) <https://arxiv.org/abs/2407.07300> (forthcoming); M Anderljung et al., "Frontier AI Regulation: Managing Emerging Risks to Public Safety" (arXiv, 2023) <http://arxiv.org/abs/2307.03718>; R Bommasani et al., "Considerations for Governing Open Foundation Models" (2024) 386 Science 151 <https://doi.org/10.1126/science.adp1848>.

[19] E.g. T Davidson et al., "AI Capabilities Can Be Significantly Improved without Expensive Retraining" (arXiv, 2023) <http://arxiv.org/abs/2312.07413>; T Schick et al., "Toolformer: Language Models Can Teach Themselves to Use Tools" (arXiv, 2023)

compromise their safety features.[20] This has already led some scholars to suggest that additional measures may be needed to address the risks from downstream modifications.[21] However, research into how policymakers might do this, and the trade-offs associated with different policy approaches, has been neglected. This article seeks to address this gap by answering the following two research questions:

    (i) What regulatory challenges do downstream developers pose?
    (ii) How might policymakers address risks posed by downstream developers?

The scope of this article is restricted in three ways. First, it focusses on the EU and UK, given their regulatory efforts to date and potential for international influence. Second, it assumes that policymakers will introduce regulation on upstream foundation model developers in an attempt to mitigate the risks posed by these models. Third, it considers whether regulation should also apply to downstream developers who modify these models (as opposed to those who use a model as-is for a specific purpose, which would be out of scope).

The article proceeds as follows. Section II sets out why downstream developers' ability to increase risk poses distinct challenges to policymakers. Section III sets out three broad approaches that policymakers could take in relation to downstream developers: (i) introduce regulation directly on downstream developers ("regulate downstream"), (ii) introduce regulation on upstream developers to mitigate downstream modification risks ("regulate upstream"), and

---

<http://arxiv.org/abs/2302.04761>; R Nakano et al., "WebGPT: Browser-Assisted Question-Answering with Human Feedback" (arXiv, 2022) <http://arxiv.org/abs/2112.09332>.

[20] E.g. C Anil et al., "Many-Shot Jailbreaking" (Advances in Neural Information Processing Systems, San Diego, 2024) <https://perma.cc/H9L3-ALAP>; X Qi et al., "Fine-Tuning Aligned Language Models Compromises Safety, Even When Users Do Not Intend To!" (arXiv, 2023) <http://arxiv.org/abs/2310.03693>; P Gade et al., "BadLlama: Cheaply Removing Safety Fine-Tuning from Llama 2-Chat 13B" (arXiv, 2024) <http://arxiv.org/abs/2311.00117>; X Yang et al., "Shadow Alignment: The Ease of Subverting Safely-Aligned Language Models" (arXiv, 2023) <http://arxiv.org/abs/2310.02949>; D Halawi et al., "Covert Malicious Finetuning: Challenges in Safeguarding LLM Adaptation" (arXiv, 2024) <http://arxiv.org/abs/2406.20053>; D Kang et al., "Exploiting Programmatic Behavior of LLMs: Dual-Use Through Standard Security Attacks" (arXiv, 2023) <http://arxiv.org/abs/2302.05733>.

[21] P Henderson et al., "Safety Risks from Customizing Foundation Models via Fine-Tuning" (Stanford University 2024) <https://perma.cc/TNC3-F97J>; D Carpenter and C Ezell, "An FDA for AI? Pitfalls and Plausibility of Approval Regulation for Frontier Artificial Intelligence" (AAAI/ACM Conference on AI, Ethics, and Society, San Jose, 2024) <https://doi.org/10.1609/aies.v7i1.31633>; M Stein and C Dunlop, "Safe before Sale" (Ada Lovelace Institute 2023) <https://perma.cc/Z4F7-R9HQ>; M Srikumar, J Chang and K Chmielinski, "Risk Mitigation Strategies for the Open Foundation Model Value Chain" (Partnership on AI 2024) <https://perma.cc/W5SA-PR9R>; S Kapoor et al., "On the Societal Impact of Open Foundation Models" (arXiv, 2024) <http://arxiv.org/abs/2403.07918>.

(iii) do not impose any new regulations in relation to downstream modification risks, but clarify existing rules or issue voluntary guidance. Section IV sets out policy recommendations. Section V concludes with suggestions for further research.

## II. What regulatory challenges do downstream developers pose?

In this section, we explore the extent to which downstream developers pose regulatory challenges. Specifically, we discuss the key role that downstream developers play in the value chain (Section II.1), the ways downstream developers can increase risk (Section II.2), and why this presents distinct challenges for policymakers (Section II.3).

*1. Downstream developers play a key role in the value chain*

The AI value chain is complex and involves many actors. Below, we set out the value chain and the role of downstream developers within it. Table 1 provides an overview of the different actors. Figure 1 illustrates how these actors relate to each other.

*Table 1*: Overview of actors along the AI value chain[22]

| Name | Description | Examples |
| --- | --- | --- |
| Compute and cloud providers | Provide computing power and infrastructure to train and run foundation models | Nvidia, Amazon Web Services, Google Cloud, IBM Cloud, Microsoft Azure, Coreweave |
| Data providers | Provide datasets for model training and refinement | Common Crawl, Scale AI |
| Foundation model developers | Design and train models using data and compute | Meta (Llama), OpenAI (GPT), Anthropic (Claude), Google DeepMind (Gemini) |
| Model hubs and hosting providers | Make foundation models available to downstream developers | GitHub, Hugging Face, Amazon Bedrock, Microsoft Azure |

---

[22] Adapted from M Srikumar, J Chang and K Chmielinski, "Risk Mitigation Strategies for the Open Foundation Model Value Chain" (Partnership on AI 2024) <https://perma.cc/W5SA-PR9R>.

| Downstream developers | Fine-tune or otherwise modify foundation models | Meditron, Vicuna, Agentforce, individual developers |
| Distribution platforms | Provide platforms where applications based on foundation models can be shared | Apple App Store, OpenAI ChatGPT Store |
| Users | Use services or applications, either as a business (B2B) or a consumer (B2C) | Finance companies, chatbot users, medical patients |

*Overview of the AI value chain*. The AI value chain involves multiple actors which interact in various ways, as illustrated in Figure 1. Upstream foundation model developers obtain compute, cloud storage, and data from a range of sources to develop their models. Downstream developers then access these models – either through a model hub or hosting provider, or directly via upstream developers (e.g. OpenAI's API). End users, in turn, can access the models directly from the upstream developer, or indirectly through downstream developers or distribution platforms. Notably, one actor can fulfil multiple roles – for example, a foundation model developer may also serve as a distribution platform (e.g. OpenAI offers GPT models and operates the GPT Store).[23] Although regulation could target a variety of actors,[24] this article focusses specifically on downstream developers.

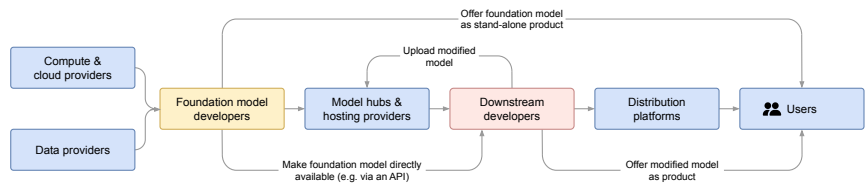

*Figure 1*: Overview of the AI value chain

*Defining "downstream developer"*. There is no generally accepted definition of the term "downstream developer". For the purposes of this article, we define downstream developer as any actor that fine-tunes or makes other modifications to a foundation model that was developed by another actor. "Fine-tuning" refers to further training after pre-training – typically using a much smaller dataset – often to elicit specific capabilities from the model.[25] "Other modifications" refers to other techniques to alter the model's internals (i.e. the model's weights) or externals (e.g. through the use of scaffolding). We provide some examples of downstream modifications below.[26] Our definition is intentionally broad to capture the full range of actors in this segment of the value chain. We discuss how this definition could be narrowed for use in a regulatory context below.[27]

---

[23] An upstream developer may also make post-training modifications to its own model. However, this article focusses on downstream developers who would not be in scope of upstream regulation.

[24] On the role of compute providers in AI governance, see e.g. L Heim et al., "Governing Through the Cloud: The Intermediary Role of Compute Providers in AI Regulation" (arXiv, 2024) <https://arxiv.org/abs/2403.08501>.

*Number of downstream developers*. Downstream developers represent a large and growing segment of the AI value chain. The number of downstream developers appears to be in the thousands, with the Hugging Face Hub alone reportedly hosting over 900,000 models, many of which seem to be modified versions of pre-existing models.[28] One key reason why this segment is so large is the relatively low barrier to entry: downstream developers require far less capital than upstream developers. Their number seems to be growing rapidly. For example, Microsoft's Azure OpenAI service, which permits downstream modifications to models, grew from 1,000+ organisations in early 2023[29] to 18,000+ by late 2023.[30] This upward trajectory is likely driven by continued development of user-friendly tools, growing demand (and funding) for bespoke models, and an ever-widening range of applications. As foundation models become more capable, and our understanding of them deepens, we should expect this trend to continue.

*Composition of downstream developers*. Downstream developers are varied in nature. They include individuals, academic researchers, start-ups, small and medium-sized enterprises, and multinational corporations. Their expertise varies by domain. While some specialise in finance, healthcare, or cybersecurity, others develop more general-purpose AI tools. This diverse community also includes malicious actors who may deliberately seek to increase risk. Because their motivations, resources, and objectives differ widely, downstream developers play several roles in shaping the AI ecosystem.

*Role of downstream developers*. Downstream developers play at least three important roles in the AI value chain. First, they can adapt foundation models for specific domains, tasks, and applications.[31] Upstream developers are unlikely to do this themselves because of limited capacity and a lack of sector-specific knowledge. Second, downstream developers may improve a foundation model more generally – for example, by combining post-training modification techniques in different ways.[32] These techniques may become even more important for AI progress if current scaling laws in pre-training start to show diminishing returns, as some have suggested.[33] Finally, they can contribute to improvements in AI safety by identifying safety issues and developing techniques to address them. For example, a cyber security firm that adopts and modifies a foundation model might make modifications to improve its cyber defence capabilities.

---

[25] T Davidson et al., "AI Capabilities Can Be Significantly Improved without Expensive Retraining" (arXiv, 2023) <http://arxiv.org/abs/2312.07413>.

[26] See Section II.2.

[27] See Section III.1.

[28] Hugging Face, "Hugging Face Hub Documentation" (2025) <https://perma.cc/N27R-VVCX>.

[29] E Boyd, "ChatGPT Is Now Available in Azure OpenAI Service" (*Microsoft*, 9 March 2023) <https://perma.cc/HA3B-3645>.

[30] A Beatman, "Microsoft Ignite 2023: Azure OpenAI Service Announces New Models and Multimodal Advancements" (*Microsoft*, 15 November 2023) <https://perma.cc/F834-Y8EQ>.

### 2. Downstream developers can increase risks

Downstream developers can fine-tune or otherwise modify foundation models in various ways. These modifications could amplify existing risks or create new ones. Below, we discuss how different types of modifications might increase risk, the circumstances under which this is more likely, and examples of real-world harm.

*Removing or evading safety features*. Downstream developers can increase risk by removing or evading a foundation model's safety mechanisms. One way to do this is through fine-tuning. Research has shown that current models' safety fine-tuning can be undone at a relatively low cost, while largely preserving the model's capabilities.[34] More concerning, these changes can be made covertly or even unintentionally. For example, one study demonstrated how it was possible to covertly fine-tune OpenAI's GPT-4 via an API so that it would act on harmful instructions.[35] Another study found that fine-tuning with benign and commonly used datasets on both open and API-accessible models can inadvertently degrade their safety alignment.[36] Beyond fine-tuning, other types of downstream modifications can also subvert safety features. For example, prompt engineering has been used to override safety controls in models developed by OpenAI, Anthropic, and Google DeepMind.[37] Improving the robustness of safeguards is an active research topic amongst upstream developers.[38]

---

[31] R Bommasani et al., "On the Opportunities and Risks of Foundation Models" (arXiv, 2022) <http://arxiv.org/abs/2108.07258>.

[32] T Davidson et al., "AI Capabilities Can Be Significantly Improved without Expensive Retraining" (arXiv, 2023) <http://arxiv.org/abs/2312.07413>.

[33] E.g. A Narayanan and S Kapoor, "Is AI Progress Slowing Down?" (*AI Snake Oil*, 18 December 2024) <https://perma.cc/5BN3-HHYD>; R Metz et al., "OpenAI, Google and Anthropic Are Struggling to Build More Advanced AI" (*Bloomberg*, 13 November 2024) <https://www.bloomberg.com/news/articles/2024-11-13/openai-google-and-anthropic-are-struggling-to-build-more-advanced-ai> accessed 10 March 2025.

[34] P Gade et al., "BadLlama: Cheaply Removing Safety Fine-Tuning from Llama 2-Chat 13B" (arXiv, 2024) <http://arxiv.org/abs/2311.00117>; X Yang et al., "Shadow Alignment: The Ease of Subverting Safely-Aligned Language Models" (arXiv, 2023) <http://arxiv.org/abs/2310.02949>.

[35] This involved eluding defence mechanisms, including dataset inspection and input/output classifiers. See D Halawi et al., "Covert Malicious Finetuning: Challenges in Safeguarding LLM Adaptation" (arXiv, 2024) <http://arxiv.org/abs/2406.20053>.

*Improving model capabilities*. Downstream developers can also increase risk by improving a foundation model's capabilities, either in a specific domain or more generally. Davidson et al. identify a range of techniques that can significantly improve a model's performance without expensive re-training (see Table 2). To measure these enhancements, they introduce the concept of "compute equivalent gains", which represents how much additional training compute would be needed to achieve similar improvements. Most of the post-training enhancements they studied improved benchmark performance by more than a 5x increase in training compute, with many exceeding 20x. Notably, these improvements were relatively inexpensive. For example, fine-tuning costs were typically less than 1% of the model's original training cost.

*Table 2*: Examples of downstream modifications[39]

| Name | Description | Example |
| --- | --- | --- |
| Tool use | Teaching a model to use new tools, like a web browser | Using Toolformer to fine-tune a model to use a search engine[40] |
| Prompting | Changing the text-based input to the model to steer its behaviour and reasoning | Using Chain-of-Thought prompting to get the model to think through the problem step-by-step[41] |

---

[36] X Qi et al., "Fine-Tuning Aligned Language Models Compromises Safety, Even When Users Do Not Intend To!" (arXiv, 2023) <http://arxiv.org/abs/2310.03693>.

[37] C Anil et al., "Many-Shot Jailbreaking" (*Advances in Neural Information Processing Systems*, San Diego, 2024) <https://perma.cc/H9L3-ALAP>; D Kang et al., "Exploiting Programmatic Behavior of LLMs: Dual-Use Through Standard Security Attacks" (arXiv, 2023) <http://arxiv.org/abs/2302.05733>.

[38] C Criddle, "Anthropic makes 'jailbreak' advance to stop AI models producing harmful results" (*Financial Times*, 2025) <https://perma.cc/V9J6-HM3D>.

[39] Adapted from T Davidson et al., "AI Capabilities Can Be Significantly Improved without Expensive Retraining" (arXiv, 2023) <http://arxiv.org/abs/2312.07413>.

[40] T Schick et al., "Toolformer: Language Models Can Teach Themselves to Use Tools" (arXiv, 2023) <http://arxiv.org/abs/2302.04761>.

[41] J Wei et al., "Chain-of-Thought Prompting Elicits Reasoning in Large Language Models" (arXiv, 2023) <http://arxiv.org/abs/2201.11903>.

| | | |
|---|---|---|
| Scaffolding | Using programs to structure the model's reasoning and the flow of information between different instances of a model or multiple models | Developing programs that allow a model to power an autonomous agent[42] |
| Solution choice | Using techniques to get the model to generate and then choose between multiple candidate solutions to a problem | Using a fine-tuned model as a verifier to rate candidate solutions and submit the one with the highest rating[43] |
| Data | Using techniques to generate more or higher-quality data that can be used for fine-tuning | Fine-tuning the model on self-generated data[44] |

*Modifying open models*. Post-training modifications can potentially increase risk in all types of foundation models, regardless of how they are released.[45] Model releases can range from fully closed to fully open.[46] Downstream developers can typically make more significant modifications to fully open models because they have access to the model's weights. In addition, foundation model developers are less likely to have oversight or control over downstream modifications of an open model. This may have implications for the policy approach.[47] For example, a downstream developer who has access to a model's weights may be better equipped to identify and mitigate risks from any modifications they have made, relative to a downstream developer that has fine-tuned a model via an API.

*Modifying highly capable models*. Policymakers are increasingly categorising foundation models by their level of capability. The underlying assumption is that model capabilities are a proxy for risk.[48] For example, the EU AI Act imposes more extensive obligations on general-purpose AI models if they have "high-impact capabilities",[49] while the UK plans to focus regulation on the "most powerful" models.[50] This is a sensible approach, as it targets obligations on developers of models that pose the most significant risks. The same principle should arguably also apply to downstream modifications, because modifying a highly capable model may be more likely to create unacceptable risks. At the same time, regulators should recognise that, in theory, even a less advanced model, if significantly enhanced by a downstream developer, may pose unacceptable risks.

---

[42] See e.g. Significant-Gravitas, "AutoGPT" (*GitHub*) <https://perma.cc/E5TK-LGHE>.

[43] K Cobbe et al., "Training Verifiers to Solve Math Word Problems" (arXiv, 2021) <http://arxiv.org/abs/2110.14168>.

[44] P Haluptzok, M Bowers and AT Kalai, "Language Models Can Teach Themselves to Program Better" (arXiv, 2023) <http://arxiv.org/abs/2207.14502>.

[45] P Henderson et al., "Safety Risks from Customizing Foundation Models via Fine-Tuning" (Stanford University 2024) <https://perma.cc/TNC3-F97J>.

[46] I Solaiman, "The Gradient of Generative AI Release: Methods and Considerations" (arXiv, 2023) <https://arxiv.org/abs/2302.04844>.

[47] For more information on policy considerations for open models, see E Seger et al., "Open-Sourcing Highly Capable Foundation Models: An Evaluation of Risks, Benefits, and Alternative Methods for Pursuing Open-Source Objectives" (arXiv, 2023) <http://arxiv.org/abs/2311.09227>; S Kapoor et al., "On the Societal Impact of Open Foundation Models" (arXiv, 2024) <https://arxiv.org/abs/2403.07918>; R Bommasani et al., "Considerations for Governing Open Foundation Models" (2024) 386 Science 151 <https://doi.org/10.1126/science.adp1848>.

*Examples of real-world harm*. Verifiable examples of downstream developers causing concrete harm are surprisingly scarce, potentially due to the difficulty of proving a direct causal link with a modification. Media coverage also tends to focus on the misuse of AI systems, rather than the underlying model and whether it was modified. However, the potential for real-world harm is there. For example, downstream modifications to generative image models such as Stable Diffusion have facilitated the creation of more realistic deepfakes.[51] We have also seen media coverage of deepfakes being used in the creation of non-consensual pornographic images,[52] as well as to spread disinformation.[53] Although we do not know whether these deepfakes were created using modified models, it seems plausible that this was the case in at least some instances. Disclaimers on liability for downstream modifications also suggest that concerns about downstream modifications exist amongst upstream

---

[48] See L Koessler et al., "Risk Thresholds for Frontier AI" (arXiv, 2024) <https://arxiv.org/abs/2406.14713>; G Sastry et al., "Computing Power and the Governance of Artificial Intelligence" (arXiv, 2024) <https://arxiv.org/abs/2402.08797>. For an overview of dangerous model capabilities, see T Shevlane et al., "Model Evaluation for Extreme Risks" (arXiv, 2023) <http://arxiv.org/abs/2305.15324>; M Phuong et al., "Evaluating Frontier Models for Dangerous Capabilities" (arXiv, 2024) <https://arxiv.org/abs/2403.13793>.

[49] Art. 51 contains the main classification rule. The term "high-impact capabilities" is defined in Art. 3(64). Obligations are set forth in Art. 55.

[50] HM King Charles III, "The King's Speech 2024" (17 July 2024) <https://perma.cc/3LH5-7M5G>.

[51] Y Chen et al., "Text-Image Guided Diffusion Model for Generating Deepfake Celebrity Interactions" (arXiv, 2023) <http://arxiv.org/abs/2309.14751>.

[52] C Newman, "Hundreds of British Celebrities Victims of Deepfake Porn" (*Channel 4*, 21 March 2024) <https://perma.cc/YAF6-BGXP>.

[53] E.g. P Glynn, "Scarlett Johansson Warns of 'AI Misuse' After Fake Kanye Video" (*BBC*, 13 February 2025) <https://www.bbc.co.uk/news/articles/c0qwkdlxgxno> accessed 10 March 2025; M Spring, "Labour's Wes Streeting Among Victims of Deepfake Smear Network on X" (*BBC*, 7 June 2024) <https://www.bbc.co.uk/news/articles/cg33x9jm02ko> accessed 10 March 2025.

developers.⁵⁴ As the number of downstream developers continues to grow, we should expect to see more harms like these materialise.

Downstream developers are able to modify foundation models in ways that increase risk, and it seems likely this will increase in future. As industry self-regulation may be insufficient,⁵⁵ policy interventions to address the risks from downstream modifications should be explored.

*3. Downstream developers present distinct challenges to policymakers*

Despite the potential need to address the risks posed by downstream developers, doing so poses a number of challenges. Below, we set out these challenges.

*Downstream developers can undermine regulatory efforts upstream*. Downstream developers could modify foundation models in ways that undermine emerging regulatory efforts upstream. For example, a downstream developer might make a modification that evades safeguards that an upstream developer has put in place to meet their regulatory obligations.⁵⁶

*Downstream developers may have limited visibility and control*. Upstream and downstream developers have control over different aspects of a model, resulting in "distributed responsibility" along the AI value chain.⁵⁷ Since downstream developers may only have partial knowledge of how the underlying model was developed and limited access to it, it may be difficult for them to identify risks (e.g. if they unknowingly degrade a model's safeguards) and mitigate them (e.g. if they only have API access). Inversely, it is unlikely that upstream developers will be able to predict or address risks stemming from all potential downstream modifications to their model.

*Downstream developers are numerous and diverse in nature*. Given downstream developers represent such a large and diverse segment,⁵⁸ a one-size-fits-all policy approach is unlikely to be effective. Therefore, policymakers may need to combine several policy tools to effectively address downstream modification risks.

---

⁵⁴ See e.g. OpenAI, "Business Terms" (14 November 2023) <https://openai.com/policies/business-terms> accessed 10 March 2025.

⁵⁵ J Schuett et al., "From Principles to Rules: A Regulatory Approach for Frontier AI" in P Hacker, A Engel, S Hammer and B Mittelstadt (eds), *The Oxford Handbook on the Foundations and Regulation of Generative AI* (Oxford University Press) <https://arxiv.org/abs/2407.07300> (forthcoming); H Toner and T McCauley, "AI Firms Mustn't Govern Themselves, Say Ex-Members of OpenAI's Board" (*The Economist*, 26 May 2024) <https://perma.cc/RY6L-RHPX>.

⁵⁶ See Section II.2.

⁵⁷ J Cobbe, M Veale and J Singh, "Understanding Accountability in Algorithmic Supply Chains," (ACM Conference on Fairness, Accountability, and Transparency, Chicago, 2023) <https://doi.org/10.1145/3593013.3594073>.

⁵⁸ See Section II.1.

*Downstream developers have considerable economic value.* The role of downstream developers (as discussed in Section II.1) means that they are important for innovation and may play a role in achieving governments' economic ambitions. Therefore, any intervention (either upstream or downstream) would need to be carefully designed to ensure that it is not overly burdensome or restrictive.

Fortunately, policymakers have a range of options for addressing the risks posed by downstream developers. We turn to these next.

## III. How might policymakers address risks posed by downstream developers?

There are three broad approaches that policymakers could take with regards to downstream developers. First, they could impose regulations directly on downstream developers ("regulate downstream") (Section III.1). Second, they could require upstream developers to implement measures that mitigate downstream modification risks ("regulate upstream") (Section III.2). Finally, they could use alternative policy tools, such as clarifying how tort law applies or issuing voluntary guidance (Section III.3). The following section describes each approach in more detail and discusses their respective advantages and disadvantages, with an overview provided in Table 3. It is important to note that these approaches are not mutually exclusive and that they may be best used in combination. For broader context, Appendix A offers examples of regulatory mechanisms used in other high-stakes industries with complex value chains.

*Table 3*: Approaches policymakers could take with regards to risks from downstream modifications

| Options | Advantages | Disadvantages |
| --- | --- | --- |
| Introduce regulation directly on downstream developers | <ul><li>Enables direct intervention by the regulator against risky downstream modifications</li><li>Clarifies downstream developers' responsibilities</li><li>Can improve regulators' understanding of downstream developers and the risks they present</li></ul> | <ul><li>Increases the scope of the regime, which could create a number of challenges (e.g. the potential to stifle to innovation)</li><li>Downstream developers may be limited in their ability to identify and mitigate risks if they do not have access to the underlying model or sufficient information about it</li></ul> |
| Introduce regulation on upstream developers to mitigate downstream modification risks | <ul><li>Helps to indirectly mitigate risk for some types of models and modifications</li><li>Could formalise and build on existing efforts by upstream developers</li><li>Keeps the scope of the regulatory regime narrow</li></ul> | <ul><li>Direct intervention by the regulator against risky downstream modifications may not be possible</li><li>Potentially stifling to innovation (e.g. if downstream modifications are overly restricted)</li><li>Upstream developers are unlikely to be able to predict and guard against all potential downstream modification risks</li></ul> |
| Do not impose any new regulations in relation to downstream modification risks, but clarify tort law or issue voluntary guidance | Tort law:<ul><li>Keeps the scope of the regulatory regime narrow</li><li>Can address unknown risks that may be difficult to specify in regulation</li><li>Limited to realised risks (rather than speculative risks)</li></ul>Voluntary guidance:<ul><li>Keeps the scope of the regulatory regime narrow</li><li>Avoids the need to pass new legislation</li><li>Reduces the risk of regulatory flight</li></ul> | Tort law:<ul><li>Direct intervention by the regulator against risky downstream modifications may not be possible</li><li>May require significant adaption</li><li>May not sufficiently incentivise risk mitigation</li></ul>Voluntary guidance:<ul><li>Direct intervention by the regulator against risky downstream modifications may not be possible</li><li>May not sufficiently incentivise risk mitigation</li></ul> |

*1. Introduce regulation directly on downstream developers*

One option for addressing the risks from downstream modifications is to introduce obligations directly on downstream developers. Note that the EU AI Act already acknowledges that general-purpose AI models can be modified into new models.[59] This implies that downstream developers could be subject to some obligations. However, more clarity is needed about the circumstances under which this would be the case. The EU AI Office has indicated that they expect very few downstream developers to be in scope.

*How could the scope of downstream regulation be narrowed?* Ideally, downstream developers should only be subject to regulation if two conditions are met. First, they have made – or are sufficiently likely to have made – modifications that either increase existing risks to an unacceptable level or introduce new, unacceptable risks. Second, imposing obligations on the downstream developer (in addition to the upstream developer) would help to meaningfully reduce these risks. To ensure that only a subset of downstream developers are in scope, policymakers could apply criteria.[60] These criteria may be specific (e.g. "a downstream developer is in scope only if they have used more than X amount of compute to modify the model") or general (e.g. "a downstream developer is in scope only if they have reason to believe they have made modifications that are likely to increase risk to an unacceptable level"). Table 4 presents examples of specific criteria across five dimensions. While general criteria may better align with the regulation's overall objectives and would likely require fewer updates as post-training modification techniques evolve, they can also create greater uncertainty regarding who is subject to regulation. We propose one possible approach which would combine a set of specific and general criteria in Section IV.

---

[59] Recital 97.

*Table 4*: Examples of specific criteria that could be used to target a subset of downstream developers

| Dimension | Examples |
| --- | --- |
| Characteristics of the downstream developer | Downstream developers are in scope only if they have more than X employees |
| Characteristics of the modification | Downstream developers are in scope only if they have used more than X amount of compute to modify the model |
| Characteristics of the modified model | Downstream developers are in scope only if the conditions under which safety cases[61] were developed no longer hold for the modified model |

---

[60] Any criteria would ideally need to be (i) a good proxy for risk, (ii) easy to measure, (iii) difficult to game, and (iv) not overly inclusive. For a more extensive list, along with the trade-offs associated with each criterion, see Appendix B.

[61] A safety case is a structured argument, supported by evidence, that a system is safe enough in a given operational context. For more information, see MD Buhl et al., "Safety Cases for Frontier AI" (arXiv, 2024) <http://arxiv.org/abs/2410.21572>; G Irving, "Safety Cases at AISI" (AI Security Institute 2024) <https://perma.cc/MDL8-QRYU>; A Goemans et al., "Safety Case Template for Frontier AI: A Cyber Inability Argument" (arXiv, 2024) <https://arxiv.org/abs/2411.08088>; Korbak et al., "A Sketch of an AI Control Safety Case" (arXiv, 2025) <https://arxiv.org/abs/2501.17315>; B Hilton et al., "Safety Cases: A Scalable Approach to Frontier AI Safety" (AI Security Institute 2025) <https://perma.cc/UXR5-7PJB>.

| | |
|---|---|
| Characteristics of the original model | Downstream developers are in scope only if they modify a model that has been trained using more than X amount of compute |
| Characteristics of the upstream developer | Downstream developers are in scope only if the upstream developer permits "significant" modifications (e.g. fine-tuning above X data points) |

*What obligations could apply?* Obligations for downstream developers could range from high-level principles to more specific rules and requirements.[62] Principles for downstream developers could mirror those for upstream developers, but their implementation may differ. For instance, if a principle states that developers must assess and mitigate unacceptable risks arising from their models, one way an upstream developer might fulfil this is through extensive testing and evaluation of their model, including the production of safety cases. In contrast, a downstream developer might be able to meet the same principle by ensuring that the modified model remains within the bounds of the safety case produced by the upstream developer. More specific requirements could include obligations to avoid certain "risk-producing" activities, such as fine-tuning a model on data related to chemical, biological, radiological and nuclear (CBRN) materials. They could also include obligations to undertake certain "risk-reducing" activities, such as those outlined in Table 5. Any obligations should presumably be proportionate to the modifications made to the model. For example, downstream developers might be required to augment technical documentation provided by the upstream developer to reflect the changes they have made, rather than creating the documentation from scratch. Policymakers would need to carefully consider this when designing regulation. For example, they may need to introduce upstream regulation to ensure that downstream developers have appropriate access to relevant documentation. The type of obligations imposed should also be guided by the government's specific policy aims. In particular, obligations to address potentially catastrophic risks from malicious use will look different to obligations to prevent accidental harm.[63]

---

[62] See J Schuett et al., "From Principles to Rules: A Regulatory Approach for Frontier AI" in P Hacker, A Engel, S Hammer and B Mittelstadt (eds), *The Oxford Handbook on the Foundations and Regulation of Generative AI* (Oxford University Press) <https://arxiv.org/abs/2407.07300> (forthcoming).

[63] For example, accidental harm may be mitigated by ensuring up-to-date information about the model and any modifications is available along the value chain, while mitigating potentially catastrophic risks from malicious use may require extensive testing and evaluation for some models.

*Table 5*: Types of obligations that could apply to downstream developers

| Type | Description | Examples |
| --- | --- | --- |
| Registration | Require downstream developers to register with the relevant authority | The downstream developer must register with the relevant authority if they plan to modify any foundation model that exceeds a defined compute threshold |
| Documentation | Require downstream developers to keep up-to-date technical documentation about the modified model | The downstream developer must update documentation (e.g. model cards) produced by the upstream developer to account for the modifications made |
| Testing & evaluation | Require downstream developers to test and evaluate the modified model | The downstream developer must produce safety cases for the modified model |
| Safeguards | Require downstream developers to implement safeguards to mitigate unacceptable risks posed by the modified model | The downstream developer must implement additional safeguards (e.g. safety fine-tuning) if the model's original safeguards have been altered |
| Reporting | Require downstream developers to report certain information to either the relevant authority, the upstream developer, or both | The downstream developer must monitor, record and report any safety incidents resulting from the modified model to the upstream developer |
| Internal policies | Require downstream developers to put in place internal policies that promote responsible development and deployment of the modified model | The downstream developer must have a policy explaining how it intends to comply with copyright law |

*Advantages*. Placing obligations directly on downstream developers offers several advantages. First, this approach empowers regulators to directly intervene if a downstream developer has made modifications that increase risk, rather than relying on the upstream developer to take action. This would help to strengthen a regime that could otherwise be undermined by downstream modifications. Second, it would provide clarity to well-meaning downstream developers about what they should do to identify and mitigate risks that could result from their modifications, which may be important in light of existing regulatory and civil liability obligations. Third, certain "light-touch" obligations, such as registration or documentation, could be used to enhance regulators' understanding of downstream developers and the risks they pose, helping to inform decisions about whether more intensive regulations are needed. It may also increase consumer trust and uptake of these models.

*Disadvantages.* Regulating downstream developers raises two significant challenges. First, it broadens the regulatory regime to include a larger and more diverse set of actors. Such expansion could stifle innovation and increase market concentration, because regulatory burdens often have a disproportionate impact on smaller players, such as start-ups. Indeed, about half of all new tech "unicorns" in 2024 were AI-related start-ups.[64] Confronted with high compliance costs, downstream developers might decide not to enter the market, or may be forced to leave it. Alternatively, they could decide not to comply with regulation if they lack the resources to do so, or if they believe the risk of enforcement to be low. This reluctance or inability to comply would create enforcement challenges – challenges that would be amplified by the need for greater resources to monitor and enforce against a larger number of actors.[65] As discussed, one strategy to manage this challenge is to focus obligations on a subset of downstream developers, though this would make it very difficult to capture all risks. Expanding the regulatory scope to include more entities also raises concerns about the potential for the quality of safety practices that require expertise to decline (e.g. producing safety cases).[66] Second, there is uncertainty about the extent to which downstream developers can identify and mitigate risks.[67] Downstream developers may unintentionally disable safety features, or lack sufficient access to the model to implement robust safety practices. Any obligations must address, or at least account for, these limitations. Collectively, these factors might deter policymakers and industry stakeholders, making any legislation difficult to pass.

---

[64] K Mamchych, "How Many AI Companies Are There in the World?" (*Ascendix Tech*, 6 November 2024) <https://ascendixtech.com/how-many-ai-companies-are-there> accessed 10 March 2025.

[65] N Almendares, MD Gilbert and R Kerley, "Enforcement Signals Under Rules and Standards" (SSRN, 2021) <https://papers.ssrn.com/abstract=3804864>.

[66] If more entities are required to conduct safety cases, this could result in: (i) lower regulatory incentives, because regulators may have fewer resources to evaluate each safety case thoroughly; and (ii) expert shortages, because there may not be enough qualified safety case experts to serve all regulated organisations.

[67] See Section II.3.

*2. Introduce regulation on upstream developers to mitigate downstream modification risks*

Another option would be to impose obligations on upstream developers to mitigate the risks from downstream modifications. We suggest some examples in Table 6. Importantly, the EU AI Act already requires providers of general-purpose AI models to take some steps to mitigate downstream harms, such as adopting proportionate transparency measures (e.g. sharing information about the model with downstream actors), and cooperating with downstream actors to enable their compliance with relevant obligations.[68]

*Table 6*: Types of obligations that could apply to upstream developers to reduce the risks from downstream modifications

| Type | Description | Examples |
| --- | --- | --- |
| Testing & evaluation | Require upstream developers to test and evaluate whether downstream modifications could increase risks and implement measures to mitigate against them | • The upstream developer must demonstrate that they have conducted model evaluations to explore the potential for risk-increasing downstream modifications<br>• The downstream developer must implement appropriate mitigations where they have identified a downstream modification that could increase risk to unacceptable levels |
| Monitoring | Require upstream developers to monitor downstream modifications (e.g. via an API) and, if any prohibited modifications are detected, report this to the regulator, revoke the downstream developers' access, or both | • The upstream developer must monitor whether the model is being fine-tuned on CBRN-specific data and revoke access if they believe this is being done<br>• The upstream developer must monitor if safeguards have been compromised (e.g. by fine-tuning) and revoke access if they believe this has been done |
| Modification restrictions | Require upstream developers to restrict what downstream modifications can be made to the model[69] | • The upstream developer must prevent fine-tuning modifications above a certain compute threshold (e.g. 10% of the |

---

[68] See Art. 53 and 55. See also Recital 85 and 101.
[69] Modification restrictions could potentially be combined with monitoring.

| | | |
|---|---|---|
| | | compute originally used to train the model) or conduct additional checks if they do permit this |
| Know-Your-Customer (KYC) screening / Vetting | Require upstream developers to vet downstream developers before granting modification rights to the model or permitting certain types of modifications (e.g. fine-tuning using more than a certain amount of compute) | • The upstream developer must not provide access if there is evidence that suggests that the downstream developer has previously modified a model in a way that has breached obligations[70]<br>• The upstream developer must not provide access if the downstream developer is based in a rogue state |
| Buffers | Require upstream developers to include "buffers" when implementing safety requirements (e.g. producing safety cases) to account for likely downstream modifications | • The upstream developer must adhere to an enhanced set of safety practices and increase their "buffers" if they intend to permit certain types of downstream modifications |
| Licenses and agreements | Require upstream developers to specify prohibited modifications or "safe parameters" in their licenses and agreements (supported by model cards) and to enforce them[71] | • The upstream developer must specify modifications that are likely to increase risk to unacceptable levels and prohibit them (e.g. fine-tuning that would degrade safeguards)[72]<br>• The upstream developer must specify at what point a downstream developer is likely to be outside of any "buffers" |
| Information sharing | Require upstream developers to share information about the model with downstream developers | • The upstream developer must ensure information about the capabilities and limitations of the model is readily available |

---

[70] This would require a central register of some sort.

[71] This would be similar to existing requirements on social media platforms in the EU, see e.g. D Cooper and AO de Meneses, "Digital Services Act's Impact on Terms of Service" (*Covington*, 20 December 2023) <https://perma.cc/5CQA-HS6D>.

[72] This could be paired with "contracting", making it the case that the downstream developer is liable for any fines or such that the developer has to pay. See e.g. D Contractor et al., "Behavioral Use Licensing for Responsible AI" (ACM Conference on Fairness, Accountability, and Transparency, Seoul, 2022) <https://doi.org/10.1145/3531146.3533143>.

*Advantages*. Introducing obligations on upstream developers offers three main benefits. First, obligations can help to indirectly mitigate risks from downstream modifications, for example, by ensuring that safeguards remain intact (e.g. through monitoring and modification restrictions), preventing bad actors from modifying models (e.g. by conducting KYC screenings), reducing the likelihood that modifications will increase risk (e.g. by including "buffers" in safety cases), and ensuring downstream developers know what modifications are likely to increase risk (e.g. by specifying them in licences and agreements).[73] Second, it builds on work already underway by several upstream developers. For example, many upstream developers already claim to monitor downstream activity (e.g. to detect violations of their terms),[74] restrict modifications (e.g. with fine-tuning rate limits),[75] include buffers in their capability assessments,[76] experiment with post-training modifications to investigate whether they might increase risk (e.g. agent evaluations to examine how models perform with agent scaffolding),[77] and include information on what modifications are permitted in their terms and licence agreements (e.g. by forbidding any attempts to circumvent safety features).[78] Mandating these practices would ensure that all upstream actors are taking precautions. They may also be easier for downstream developers to adhere to if they were standardised. Third, this approach would avoid expanding the scope of the regime, which is undesirable for the reasons set out above.[79]

*Disadvantages*. There are, however, three significant downsides to this approach. First, it may leave the regulator unable to directly intervene if a downstream developer had modified a model which increased risk to an unacceptable level. Without such intervention, downstream developers could make harmful modifications without being subject to regulatory enforcement. This situation also introduces the potential for regulatory arbitrage. For instance, an upstream developer could establish a separate subsidiary to fine-tune a model on prohibited datasets if only the primary business were restricted from training on those types of data.[80] Second, this approach could stifle innovation and entrench upstream power. For example, if upstream developers are incentivised to prohibit downstream modifications outright, this may result in an unfavourable "misuse-use trade-off".[81] It may also encourage widespread surveillance, which would raise important concerns about data privacy, particularly if extensive monitoring, such as storing chat logs for prolonged periods, becomes necessary for ensuring compliance. Third, there is uncertainty over its effectiveness. Many of the potential upstream obligations identified would be impractical for open models.[82] Even in the case of closed models, upstream developers are unlikely to be able to manage all potentially risky downstream modifications given the difficulty predicting and analysing all possible modifications. Furthermore, downstream developers may still find ways to circumvent any controls (e.g. through covert fine-tuning).[83]

---

[73] The specific risks that can be mitigated will depend on the obligations imposed. For example, a requirement to share information may help mitigate risks from modifications that could result in accidental harm, whereas a requirement to conduct monitoring could help to also mitigate against modifications designed to intentionally increase risk.

[74] E.g. Anthropic, "The Claude 3 Model Family: Opus, Sonnet, Haiku" (4 March 2024) <https://perma.cc/XQ7S-ED9C>.

[75] E.g. OpenAI, "Fine-Tuning" <https://platform.openai.com/docs/guides/fine-tuning> accessed 10 March 2025.

[76] E.g. Google DeepMind, "Frontier Safety Framework (Version 2.0)" (4 February 2025) <https://perma.cc/28ZW-D8K4>.

[77] E.g. Anthropic, "Responsible Scaling Policy" (15 October 2024) <https://perma.cc/5GK4-W3KH>.

[78] E.g. OpenAI, "Terms of Use" (11 December 2024) <https://openai.com/policies/row-terms-of-use> accessed 10 March 2025.

[79] See Section III.1.

*3. Do not impose any new regulations in relation to downstream modification risks*

A third option would be to refrain from imposing any new regulations on downstream developers or on upstream developers for actions taken downstream. Instead, policymakers could clarify how tort law applies to downstream developers or issue voluntary guidance to help manage downstream modification risks.

*Tort law*. Policymakers could clarify or alter the application of tort with respect to downstream developers. In principle, tort law imposes a general duty on everybody to take reasonable care to avoid causing harm to persons or property. Previous work has discussed how US tort law might fill a gap in the governance of upstream developers before new regulation is enacted, and serve as a valuable auxiliary mechanism once regulation is in place.[84] It is likely that this will also be the case for UK and EU jurisdictions (at least to some extent). This rationale also applies to downstream developers, particularly because the risks they introduce may be more readily foreseeable (e.g. if the modifications are domain-specific). Importantly, any contract between the upstream and downstream developer can impact which of them bears the risk of liability. For example, the terms of service could include an indemnity clause in either party's favour. There are also ways policymakers might adapt tort law in this context (e.g. by creating an evidentiary presumption of responsibility). However, a detailed analysis of tort law and such reforms is beyond the scope of this article.

---

[80] Note that there might be ways to manage this. For example, if the entity that ultimately holds the economic authority over the development or deployment process is responsible under the regulatory regime, and not the legal person that is directly acting.

[81] M Anderljung and J Hazell "Protecting Society from AI Misuse: When are Restrictions on Capabilities Warranted?" (2024) AI & Society <https://doi.org/10.1007/s00146-024-02130-8>.

[82] Despite this, combining pre-emptive safety testing by the upstream developer with clear operating parameters for downstream developers could still help to mitigate downstream modification risks.

[83] It may be possible to partly counter this problem by restricting undesired access to a closed model after modifications are made (e.g. by removing the respective API key).

[84] M van der Merwe, K Ramakrishnan and M Anderljung, "Tort Law and Frontier AI Governance" (*Lawfare*, 24 May 2024) <https://perma.cc/9MTC-DYJX>; K Ramakrishnan, G Smith and C Downey, "U.S. Tort Liability for Large-Scale Artificial Intelligence

*Voluntary guidance*. Another option is to introduce voluntary guidance for downstream developers, upstream developers, or both, regarding downstream modification risks.[85] Guidance would not need to be legally binding (similar to the Frontier AI Safety Commitments).[86] Guidance could draw on existing frameworks, such as procurement guidance developed by the US Office of Management and Budget (OMB).[87] Guidance could serve as a foundation for future regulation or prove sufficient without new regulatory obligations. In addition, government-issued guidance might affect how courts assess whether a developer exercised reasonable care, thus interacting with tort law.

*Advantages*. These non-regulatory approaches offer several potential advantages. Both avoid expanding the regulatory regime beyond upstream developers, which may be desirable for the reasons discussed above.[88] There are several further benefits in relation to voluntary guidance. First, it may appeal to policymakers because it avoids the need to pass new legislation.[89] Relatedly, thoughtfully designed guidelines may be attractive to downstream developers if, for example, they enable them to press for risk-relevant information from upstream developers. Additionally, voluntary guidance can serve as a trial mechanism, allowing policymakers to test potential requirements before mandating them, a strategy that could help reduce the risk of regulatory flight. Tort law, on the other hand, has the advantage of being able to address currently unknown risks without requiring policymakers to specify rules or guidance in

---

Damages: A Primer for Developers and Policymakers" (RAND 2024) <https://doi.org/10.7249/RRA3084-1>.

[85] Governments could also encourage model hubs and hosting providers to adopt guidance for governing downstream developers who access models through their services. However, this is beyond the scope of this article.

[86] DSIT, "Frontier AI Safety Commitments, AI Seoul Summit 2024" (7 February 2025) <https://perma.cc/B8RH-NVRN>.

[87] SD Young, "Advancing the Responsible Acquisition of Artificial Intelligence in Government" (*OMB*, 24 September, 2024) <https://perma.cc/CL8T-X3WR>.

[88] See Section III.1.

[89] But note that policymakers may need new legislation to clarify or amend how tort law applies.

advance. Furthermore, because it only imposes liability after harm has occurred, it only targets risks that are realised (versus those that are speculative). This feature reduces the risk of overregulation and may also reduce compliance costs for downstream developers.

*Disadvantages*. Despite certain advantages, these alternative approaches may not fully address the risks from downstream modifications. Encouraging widespread adoption of voluntary guidance could be particularly challenging given the sheer number and rapid growth of downstream developers, many of whom have limited resources. Tort law also has several limitations, including the asymmetry of information between plaintiffs and developers regarding the use of foundation models and how they operate. This imbalance could make it difficult for plaintiffs to prove that a developer (either upstream or downstream) failed to take reasonable care and that this failure caused harm. Moreover, tort doctrine tends to develop slowly, potentially undermining its ability to incentivise developers to prioritise risk mitigation. Consequently, tort law may require significant adaptation to effectively manage the risks arising from downstream modifications, and even then, may fall short given the complexity of the AI value chain.

## IV. Policy recommendations

In this section, we make high-level policy recommendations. In short, we argue that policymakers should develop voluntary guidance to help mitigate downstream modification risks (Section IV.1), introduce regulation on upstream developers to mitigate downstream modification risks (Section IV.2), and monitor the downstream developer ecosystem to inform whether downstream regulation is warranted (Section IV.3).

### 1. Develop voluntary guidance to help mitigate downstream modification risks

Given the potential for downstream modifications to increase risk, policymakers should, at the very least, develop voluntary guidance for upstream and downstream developers to help mitigate risks from downstream modifications. This guidance could help establish best practice and form the basis of future regulation.

*Voluntary guidance for downstream developers*. Voluntary guidance for downstream developers could include the following recommendations: (i) Only modify foundation models whose developers have established responsible use policies that facilitate effective risk assessment and mitigation, and adhere to these policies. (ii) Only modify foundation models from developers that enable incident reporting, and promptly report any incidents arising from modified versions of the model. (iii) Only modify foundation models from developers that provide technical documentation, and update it to reflect any modifications. (iv) Whenever possible, refrain from modifying foundation models in ways that are likely to increase risk. (v) If modifications are suspected to have increased risk, conduct an assessment, potentially in coordination with the upstream developer, and only proceed with deployment if these risks can be reduced to an acceptable level.

*Voluntary guidance for upstream developers*. Voluntary guidance for upstream developers could include the following recommendations: (i) Provide downstream developers with relevant information so they can better understand whether a modification is likely to increase risk. (ii) Establish responsible use policies that provide assurance to downstream developers that their modifications are safe (e.g. by clearly setting out any prohibited modifications). (iii) Wherever possible, notify downstream developers if their modifications appear likely to have increased risk. (iv) Invest in research aimed at enhancing the robustness of safeguards to prevent risk-increasing modifications. (v) Encourage downstream developers to report any incidents that arise from modified versions of the model. (vi) Provide clear versioning practices to improve traceability for downstream developers and other stakeholders (e.g. regulators). (vii) Provide or recommend evaluation benchmarks and other methodologies for testing the modified model. (viii) Collaborate with policymakers to share insights on which modifications have the greatest potential to increase risks.

*2. Introduce regulation on upstream developers to mitigate downstream modification risks*

If and when policymakers introduce regulation requiring upstream developers to mitigate risks posed by their models, they should include requirements to address risks from potential downstream modifications. In particular, policymakers should clarify that upstream developers must take into account how downstream developers might modify their models when conducting risk assessments. Policymakers could impose a duty on upstream developers to determine which modifications are likely to increase risks to unacceptable levels and to implement safeguards where feasible and appropriate. While responsibility to determine which modifications pose such risks would rest with the upstream developer, relevant authorities could provide guidance on identifying and evaluating them. Additionally, policymakers could specify practices to fulfil this duty, such as thoroughly ensuring the model's capacity for harmful capabilities is sufficiently explored, monitoring the effectiveness of existing safeguards, and setting safe operating parameters.[90] Some flexibility in these measures will be necessary, because the measures available to upstream developers vary depending on factors like the method of release. Given the rapid pace of research and innovation, there is also an inherent uncertainty about how effective these measures will prove in practice, so policymakers should monitor their impact.

---

[90] For more examples of specific practices, see Table 6.

*3. Monitor the downstream developer ecosystem to inform whether downstream regulation is needed*

While obligations on upstream developers may significantly help to mitigate the risks from downstream modifications, we expect that a subset of downstream developers may eventually require certain obligations as well. Before introducing such requirements, we recommend that policymakers closely monitor instances of harm involving foundation models and investigate whether those harms resulted from modified versions. If sufficient harms arise from modified models, despite voluntary guidance and upstream obligations, this would strengthen the case for downstream regulation. Nonetheless, it is paramount that any obligations are targeted and proportionate to avoid unacceptable impacts on the downstream developer ecosystem.

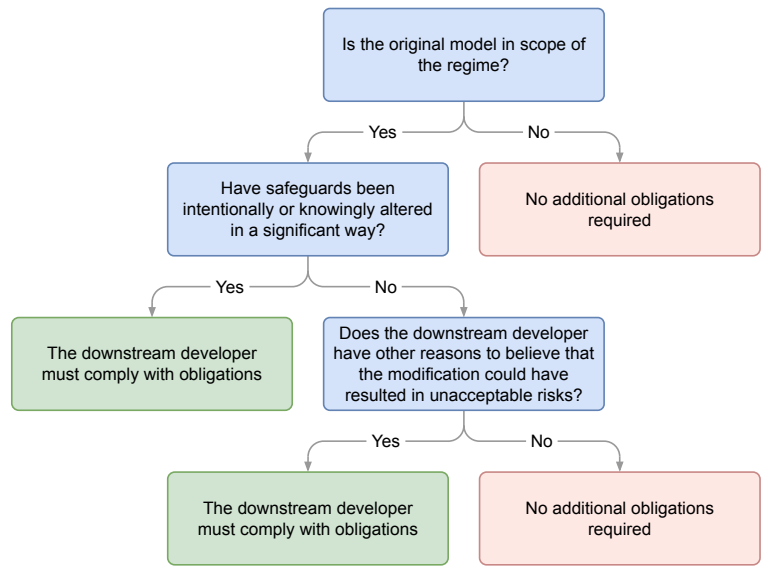

*Figure 2*: One possible approach for identifying downstream developers whose modifications might warrant regulatory intervention

One possible approach is illustrated in Figure 2. The goal of this approach is to impose obligations on downstream developers only when their modifications could have resulted in an unacceptable level of risk. This would be challenging to specify in regulation, but we think the suggested combination of criteria could strike a balance between capturing modifications that sufficiently increase risk, without being overly inclusive.

*Is the original model in scope of the regime?* Downstream developers would only have the potential to be in scope of obligations if they modify a model that is already regulated. This reflects the principle that obligations on downstream developers aim to reinforce the regulatory framework, rather than expand it to models that would not otherwise be subject to oversight.

*Have safeguards been intentionally or knowingly altered in a significant way?* Any downstream developer that has significantly altered the safeguards of a model in scope of the regime – either intentionally or knowingly – would be subject to obligations.[91] This captures the scenario where an upstream developer has implemented protections against known risks, and a downstream actor has actively circumvented or disabled them – an action typically prohibited by upstream developers' terms of use but one that regulators should be able to address directly.

*Does the downstream developer have other reasons to believe that the modification could have resulted in unacceptable risks?* Any downstream developer that has made modifications that could have increased risk to unacceptable levels would also be subject to obligations. This criterion acts as a "backstop" for capturing risk-increasing modifications that do not involve removing safeguards. Regulators could identify conditions indicating when risk may be unacceptable, such as: (i) the upstream developer explicitly warning that the modification could result in an unacceptable level of risk; (ii) the modified model being significantly more capable than the original on tasks used in high-risk domains; and (iii) the modification having used a substantial amount of compute.[92]

Under this approach, two primary obligations could apply to downstream developers in scope. First, they would be required to register with the relevant authority, providing information about the model they have modified, the modifications they made, and what they intend to do with the model. Second, they would need to demonstrate either that (i) their modifications do not increase risk to unacceptable levels (e.g. by showing that the modifications do not materially change the original's safety case), or (ii) they have implemented measures to effectively mitigate the risk (e.g. through additional safeguards or confining the model's use to a restricted environment).

---

[91] An exemption could be given if the upstream developer has shown that the model is incapable of posing unacceptable risk, even where safeguards are removed.

[92] Although additional work is needed to validate this assumption, tying obligations to high compute usage could more effectively capture significant capability enhancements. Importantly, such a threshold might initially be set very high (e.g., 10% of the compute threshold used to define which foundation models are in scope of obligations on upstream developers) to avoid expanding the regime too far.

Although additional research is needed to be confident this is the right approach, we think it balances the trade-offs outlined in this article. There are at least three reasons for this. First, it ensures that upstream developers retain accountability for downstream modifications to their models. Second, it begins with a relatively narrow scope, reducing the risk of stifling innovation or overburdening enforcement, but leaves room for gradual expansion if needed.[93] Third, by combining specific and general criteria, it targets downstream developers that are more likely to increase risk to unacceptable levels and means it would be relatively difficult to game

Nonetheless, we recognise this approach has limitations. For example, incorporating a compute threshold may pose practical challenges, as downstream developers may not always be able to accurately assess their compute usage. One alternative could be to consider the amount of data used for fine-tuning. In addition, using general criteria (e.g. whether a modification may have resulted in unacceptable risk) places responsibility on downstream developers to make this judgement. However, we view this flexibility as necessary, given the wide range of possible modifications and the rapidly evolving nature of these techniques.

## V. Conclusion

The article has made three main contributions. First, it has highlighted the importance of downstream developers as a potentially overlooked regulatory target. Second, it has evaluated policy options for mitigating downstream modification risks. Third, it has proposed a regulatory approach for mitigating downstream modification risks that avoids unnecessary burdens on downstream developers.

At the same time, several important questions remain. Further work could: (i) Provide a comprehensive mapping of the downstream developer ecosystem. (ii) Investigate how likely it is that different types of downstream modifications will increase risk – intentionally or otherwise. (iii) Analyse the potential economic impacts of regulatory interventions aimed at mitigating the risk from downstream modifications. (iv) Examine whether other actors along the value chain warrant regulatory intervention (e.g. model hubs and hosting providers). (v) Explore how regulation at different points along the entire value chain can be designed in the most streamlined way.

---

[93] See Annex B for examples.

Ultimately, policymakers need to take a holistic approach when designing AI regulations. Many current efforts focus on regulating upstream developers, but downstream developers are often overlooked. If this gap remains, the regulatory framework could be vulnerable to loopholes and fail to manage risks appropriately. Although this is a complex issue, policymakers have several options. By combining these measures, they can effectively mitigate risks without stifling downstream innovation.

# Acknowledgements

We are grateful for valuable comments and feedback from Alan Chan, Alexander Erben, Ben Bucknall, Ben Garfinkel, Christoph Winter, Connor Dunlop, Cristian Trout, Irene Solaiman, Jaime Sevilla, John Lidiard, Lauritz Morlock, Leia Wang, Lennart Heim, Leonie Koessler, Marie Buhl, Maximilian Negele, Michael Chen, Peter Wills, Sam Manning, Sayash Kapoor, Stephen Clare, and Vinay Hiremath (in alphabetical order). All remaining errors are our own.

# Appendix A: Case studies from other industries

Regulators in other high-stakes industries with complex value chains also need to manage risks that arise downstream. Below, we provide some examples of regulatory mechanisms used to mitigate downstream risks in other industries, namely pharmaceuticals (Appendix A.1), aviation (Appendix A.2), chemicals and food (Appendix A.1), and consider how they might be applied in the context of downstream AI regulation.

### A.1 Pharmaceuticals: Regulating generic drugs and drug alterations

The pharmaceutical value chain spans initial drug discovery and development, through manufacturing, distribution, and marketing. Downstream manufacturers (e.g. generic drug companies) produce pharmaceuticals based on original formulas, and even the original manufacturer of a drug may make post-approval changes (e.g. new formulations, manufacturing tweaks). These modifications carry risks: a generic that is not equivalent to the original could be less effective or unsafe, and a process change might alter the purity or potency of a drug. Regulators therefore treat any deviation from the approved medicine as potentially risky until proven otherwise.

*Approval process for generic drugs*. In the US, UK, and EU, generic drugs must go through a regulatory approval process before they can be sold. The generic manufacturer must demonstrate "equivalence" to the brand-name drug on multiple fronts.[94] For example, the US FDA requires evidence that a generic has the same active ingredient, dosage form, and strength.[95]

*Change control for approved drugs*. Regulators also control any post-approval changes to medicines. In the EU and UK, changes to a licensed drug (called "variations") are classified by risk: minor changes (like slight manufacturing tweaks or editorial label updates) can be implemented with notification, while major changes that could affect a drug's safety, quality or efficacy require prior regulatory approval.[96] A significant change in formulation triggers a review and re-approval by authorities before the modified drug can be marketed.

Before deploying a modified foundation model, the downstream developer could be required to demonstrate that it meets safety benchmarks like the original (or to an agreed standard) – this would be analogous to showing a generic has the same therapeutic effect as the brand drug. Like pharmaceuticals, a change control process could also be established for foundation models. For example, if foundation models needed regulatory approval before release, then major downstream modifications may warrant a re-approval process to confirm they do not increase risk.

### A.2 Aviation: Certification of aircraft modifications

The aviation industry's supply chain involves aircraft manufacturers, engine makers, airlines, and maintenance and repair organisations. Airplanes often serve for decades, during which time they undergo upgrades and fixes – new software in flight control systems, replacement of parts with improved versions, or retrofitting of new technology into older airframes. Any downstream modification can carry safety-critical risk.

*Type Certificates and Supplemental Type Certificates*. In the US and EU, no aircraft or major component can be used commercially without a type certificate proving it meets safety regulations. When an actor (e.g. the manufacturer or a third-party company) wants to modify an already-certified aircraft, they must obtain a Supplemental Type Certificate (STC).[97] For example, if an aviation company develops a winglet retrofit to improve fuel efficiency, an STC is required to certify that change. The STC forces the modifier to test and document the effects of the change on the aircraft's performance and safety and confirms that the proposed modification is airworthy and compatible with the original design. Notably, the STC does not just approve the new part, but how it affects the original aircraft design, to ensure the whole system remains safe.

---

[94] K Uhl and JR Peters, "How the FDA Ensures High-Quality Generic Drugs" (2018) 97 American Family Physician 696 <https://perma.cc/7YMF-8WEM>.

[95] FDA, "What's Involved in Reviewing and Approving Generic Drug Applications?" (23 April 2024) <https://perma.cc/2N3T-NFAC>.

[96] TOPRA, "Lifecycle Management: EU and US Variation Requirements" (2017) <https://perma.cc/BN2E-WQKJ>.

[97] FAA, "Supplemental Type Certificates" (7 July 2023) <https://perma.cc/2832-6UVX>; EASA, "Supplemental Type Certificates" (21 February 2024) <https://perma.cc/S6HG-3P77>.

*Tiered approach*. Aviation regulators differentiate between minor changes (which do not appreciably affect weight, balance, structural strength, reliability, or operational characteristics) and major changes. If a modification is extensive (e.g. a fundamental redesign of a flight control software), authorities may require a completely new type certificate, essentially treating it as a new aircraft design.

This example underscores the importance of evaluating the impact of a modification in context. Just as an STC ensures a new aircraft component does not destabilise the plane, an evaluation of a modified model could ensure that it is still safe for use. Another lesson is the tiered approach: minor model updates might be allowed with minimal or no oversight, whereas major updates would trigger a full re-evaluation.

### A.3 Chemicals and food: Upstream controls to prevent downstream misuse and contamination

In chemical and food supply chains, products often pass through many actors before reaching end users. Upstream producers (e.g. a chemical manufacturer or a food ingredient supplier) may not have any direct contact with end users, but improper downstream use can cause serious harms (e.g. industrial chemicals diverted to make explosives or a contaminated ingredient causing foodborne illness in food products). Because of the potential downstream risks, there are duties on upstream suppliers to help mitigate them.

*Buyer verification*. Under the EU Explosives Precursors Regulation,[98] certain chemicals that can be used to manufacture bombs (e.g. ammonium nitrate) are tightly controlled. Suppliers must verify that any purchaser has a legitimate purpose for the explosive, and obtain documentation (e.g. a license) before sale. They are also obliged to report suspicious transactions or unusually large purchases.

*Usage Instructions*. In the US food additive industry: any approved additive has strict limits for safe use, so its manufacturer must include "adequate directions for use" on the label to ensure downstream food producers stay within those safety limits.[99] This means if an actor sells an additive, they are obliged to tell the downstream manufacturer how to use it safely (e.g. concentration should not exceed X%, or store under X conditions).

---

[98] Regulation (EU) 2019/1148 On the Marketing and Use of Explosives Precursors <https://eur-lex.europa.eu/eli/reg/2019/1148/oj/eng> OJ L 186/1.

[99] E.g. 21 CFR § 172.130 Dehydroacetic Acid <https://perma.cc/KBP9-5U8E>.

*Traceability*. The EU General Food Law[100] establishes a "one step back – one step forward" traceability mandate: businesses must be able to identify who they got a food product from and to whom they supplied it.[101] If a contaminant or allergen issue is found in an ingredient, regulators can trace it upstream to the source and downstream to all final products, and the upstream supplier often must initiate a recall.

The chemical and food sectors highlight the importance of knowing one's downstream users and engaging in practices that promote safe downstream use. Just as chemical suppliers verify customers and restrict sales of dual-use substances, foundation model providers might need to implement vetting or licensing for those who want to modify their models in significant ways and report any suspicious activity. Furthermore, foundation model developers may need to set out safe operating parameters for downstream developers, akin to usage directions on food additives. AI supply chains could also borrow traceability concepts from the food industry: for example, regulators might require downstream developers to keep records of the original model and how they modified it. This would enable the regulator to trace back to the underlying model if a major incident were to occur.

---

[100] Regulation (EC) 178/2002 Laying Down the General Principles and Requirements of Food Law, Establishing the European Food Safety Authority and Laying Down Procedures in Matters of Food Safety <https://eur-lex.europa.eu/eli/reg/2002/178/oj/eng> OJ L 31.

[101] Commission Implementing Regulation (EU) No 931/2011 On the Traceability Requirements Set By Regulation (EC) 178/2002 <https://eur-lex.europa.eu/eli/reg_impl/2011/931/oj/eng> OJ L 242/2.

# Appendix B: Examples of specific criteria that could be used to target a subset of downstream developers

|  | Criteria | Example | Advantages | Disadvantages |
|---|---|---|---|---|
| Characteristics of the downstream developer | Size of downstream developer | • The downstream developer's annual revenue, market capitalisation, or number of employees exceeds X | • Targets actors with the greatest resources<br>• Relatively easy to measure | • Not a good proxy for risk<br>• Threshold would be somewhat arbitrary |
|  | Degree of access / Amount of control | • The downstream developer has control over the model's weights | • Targets actors that may be more able to mitigate risks<br>• Relatively easy to measure | • Could be seen to unfairly penalise open-source models<br>• Would not capture modifications made via APIs (but presumably the upstream developer would be more able to mitigate risk in these cases) |
| Characteristics of the modification | Cost of modifications made to the model | • The cost of the modification exceeds X (or X% of the upstream developer's costs) | • Relatively easy to measure | • Somewhat gameable (e.g. could stop modifications just below the threshold)<br>• Would not catch inexpensive modifications which may still increase risk (e.g. small amounts of fine-tuning to remove safeguards)<br>• Threshold would be somewhat arbitrary |
|  | Type of modification made to the model | • Whether the model's "internals" have been modified<br>• Whether the model has been fine-tuned<br>• Whether the model's safety features have been altered | • Some types of modifications (e.g. modifications to safety features) are a relatively good proxy for some risks<br>• Relatively easy to measure | • Not always a good proxy for risk (because e.g. it does not capture the extent of the modification)<br>• May be overly strict<br>• Not robust to developments in the techniques used to modify models<br>• If only some types of modifications are targeted (e.g. fine-tuning) other techniques may be developed to increase risk |
|  | Amount of compute used for modifications | • The amount of compute used to modify the model exceeds X (or X% of the amount originally used to train the model) | • Likely to target actors with the greatest resources (assuming they typically use more compute)<br>• Relatively easy to measure | • Would not catch low-compute modifications which may still increase risk (e.g. small amounts of fine-tuning to remove safeguards)<br>• Threshold would be somewhat arbitrary<br>• Somewhat gameable (e.g. fine-tuning could be stopped just below the threshold, or training runs could be structured to reduce the amount of compute needed) |
|  | Amount of data used for fine-tuning | • The amount of data used for fine-tuning exceeds X (or X% of the amount originally used to train the model) | • More granular, so may be a better proxy for risk than e.g. the type of modification<br>• Relatively easy to measure | • Would not catch modifications other than fine-tuning that may still increase risk<br>• Would not catch fine-tuning that uses a relatively small amount of data that may still increase risk (e.g. small amounts of fine-tuning to remove safeguards)<br>• Somewhat gameable (e.g. fine-tuning could be stopped just below the threshold)<br>• Threshold would be somewhat arbitrary |
| | Type of data used for fine-tuning | • The model has been fine-tuned on a dataset that contains CBRN data | • Targets fine-tuning that is more likely to increase some risks | • Would not catch modifications other than fine-tuning that may still increase risk<br>• May be difficult to classify data types (e.g. it may not be possible to draw a line between CBRN-related data and more general biological data) |
| Characteristics of the modified model | Degree of performance or capability improvement | • The modified model can perform new tasks or shows a marked improvement on industry benchmarks<br>• The conditions under which safety cases were developed no longer hold for the modified model | • Likely to be a good proxy for some risks | • Would require evaluation of the modified model (although it may be possible to rely on general capability benchmarks which many developers use already)<br>• Not possible to assess until after the modification has been made |
| | Increase in inference cost | • The inference cost has increased by X% compared to the original model | • Likely to be a good proxy for some risks | • Only applies to some types of modification (e.g. chain-of-thought)<br>• Clear scaling laws are not yet established<br>• Threshold would be somewhat arbitrary<br>• Not possible to assess until after the modification has been made |
| | How the modified model is released | • The modified model has been released for use in open-ended environments | • May be a predictor of how likely it is that the model will be used to cause harm | • Would not catch modified models released in other ways that may pose risks<br>• If some types of release are targeted, then others (that could also pose risk) may become more common |
| Characteristics of the original model | Amount of compute used for training | • The original foundation model was trained using more than X amount of compute | • May be a good proxy for risk (given larger models are generally more capable)<br>• Relatively easy to measure (although would rely on this information being disclosed) | • Does not capture the type or extent of modifications, so not a suitable proxy for risk on its own |
| | Foundation model size | • The size of the original foundation model exceeds X number of parameters | • May be a good proxy for risk (given larger models are generally more capable)<br>• Relatively easy to measure (although would rely on this information being disclosed) | • Does not capture the type and extent of modifications, so not a suitable proxy for risk on its own |
| | Compute-equivalent gains | • The additional training compute needed to improve the original model's benchmark performance by as much as the post-training modification exceeds X amount | • Reduces capability improvements to a single metric | • Would require evaluation of the modified model (although it may be possible to rely on general capability benchmarks which many developers use already)<br>• Difficult to measure and interpret<br>• The threshold would be somewhat arbitrary<br>• Not possible to assess until after the modification has been made |

| Characteristics of the upstream developer | Permissible modifications | • The upstream developer permits "more significant" or "risky" modifications, e.g. fine-tuning above a certain amount | • Relatively easy to measure | • Not a good proxy for risk on its own because e.g. even if a downstream developer made a minor modification, if the upstream provider permitted more extensive modifications then they would still be in scope<br>• May encourage upstream developers to highly restrict downstream modifications |
|---|---|---|---|---|
| | Upstream developer controls | • The upstream developer does not have methods to monitor and mitigate risks from downstream modifications | • Would incentivise downstream developers to use models that have controls (because they would be exempt from obligations) | • Could be seen to unfairly penalise open-source models |